\documentclass[aps,prl,twocolumn,superscriptaddress,showpacs,10pt,floatfix]{revtex4}
\usepackage{graphicx} 
\begin{document}

\title{
{\bf \boldmath 
First observation of the decay $K^+\rightarrow e^+\nu_e\mu^+\mu^-$
}
} 
\date{13 January 2006}

\affiliation{Brookhaven National Laboratory, Upton, NY 11973}
\affiliation{Department of Physics and Astronomy, University of New
Mexico, Albuquerque, NM 87131} \affiliation{Department of Physics
and Astronomy, University of Pittsburgh, Pittsburgh, PA 15260}
\affiliation{Institute for Nuclear Research of Russian Academy of
Sciences, Moscow 117 312, Russia} \affiliation{Paul Scherrer Institut,
CH-5232 Villigen, Switzerland} \affiliation{Physics Department, Yale
University, New Haven, CT 06511} \affiliation{Physik-Institut,
Universit\"at Z\"urich, CH-8057 Z\"urich, Switzerland}

\author{H.~Ma} \affiliation {Brookhaven National Laboratory, Upton, NY
11973}

\author{R.~Appel} \affiliation {Physics Department, Yale University,
New Haven, CT 06511} \affiliation{Department of Physics and
Astronomy, University of Pittsburgh, Pittsburgh, PA 15260}

\author{G.~S.~Atoyan} \affiliation{Institute for Nuclear Research of
Russian Academy of Sciences, Moscow 117 312, Russia}

\author{B.~Bassalleck} \affiliation{Department of Physics and
Astronomy, University of New Mexico, Albuquerque, NM 87131}

\author{D.~R.~Bergman} \altaffiliation[{\scalebox{0.9}[1.0]{Present
address:}}]{Rutgers University, Piscataway, NJ 08855}  \affiliation
{Physics Department, Yale University, New Haven, CT 06511}

\author{N.~Cheung} \affiliation{Department of Physics and Astronomy,
University of Pittsburgh, Pittsburgh, PA 15260}

\author{S.~Dhawan} \affiliation {Physics Department, Yale University,
New Haven, CT 06511}

\author{H.~Do} \affiliation {Physics Department, Yale University, New
Haven, CT 06511}

\author{J.~Egger} \affiliation{Paul Scherrer Institut, CH-5232
Villigen, Switzerland}

\author{S.~Eilerts} \altaffiliation[{\scalebox{0.9}[1.]{Present
address:}}]{Black Mesa Capital, Santa Fe, NM 87501}
\affiliation{Department of Physics and Astronomy, University of New
Mexico, Albuquerque, NM 87131}

\author{W.~Herold} \affiliation{Paul Scherrer Institut, CH-5232
Villigen, Switzerland}

\author{V.~V.~Issakov} \affiliation{Institute for Nuclear Research of
Russian Academy of Sciences, Moscow 117 312, Russia}

\author{H.~Kaspar} \affiliation{Paul Scherrer Institut, CH-5232
Villigen, Switzerland}

\author{D.~E.~Kraus} \affiliation{Department of Physics and Astronomy,
University of Pittsburgh, Pittsburgh, PA 15260}

\author{D.~M.~Lazarus} \affiliation {Brookhaven National Laboratory,
Upton, NY 11973}

\author{P.~Lichard} \affiliation{Department of Physics and Astronomy,
University of Pittsburgh, Pittsburgh, PA 15260}

\author{J.~Lowe} \affiliation{Department of Physics and Astronomy,
University of New Mexico, Albuquerque, NM 87131}

\author{J.~Lozano} \altaffiliation[{\scalebox{0.9}[1.]{Present
address:}}]{University of Connecticut, Storrs, CT 06269}
\affiliation {Physics Department, Yale University, New Haven, CT
06511}

\author{W.~Majid} \altaffiliation[{\scalebox{0.9}[1.]{Present
address:}}]{LIGO/Caltech, Pasadena, CA 91125}  \affiliation {Physics
Department, Yale University, New Haven, CT 06511}

\author{S.~Pislak} \altaffiliation[{\scalebox{0.9}[1.]{Present
address:}}]{Phonak AG, CH-8712 St\"afa, Switzerland.}
\affiliation{Physik-Institut, Universit\"at Z\"urich, CH-8057
Z\"urich, Switzerland} \affiliation {Physics Department, Yale
University, New Haven, CT 06511}

\author{A.~A.~Poblaguev} \affiliation{Institute for Nuclear Research
of Russian Academy of Sciences, Moscow 117 312, Russia}

\author{P.~Rehak} \affiliation {Brookhaven National Laboratory, Upton,
NY 11973}

\author{A.~Sher} \altaffiliation[{\scalebox{0.9}[1.0]{Present
address:}}]{SCIPP, University of California, Santa Cruz, CA 95064.}
\affiliation{Department of Physics and Astronomy,
University of Pittsburgh, Pittsburgh, PA 15260}

\author{J.~A.~Thompson} \altaffiliation{Deceased.} 
\affiliation{Department of Physics and
Astronomy, University of Pittsburgh, Pittsburgh, PA 15260}

\author{P.~Tru\"ol} \affiliation{Physik-Institut, Universit\"at
Z\"urich, CH-8057 Z\"urich, Switzerland} \affiliation {Physics
Department, Yale University, New Haven, CT 06511}

\author{M.~E.~Zeller} \affiliation {Physics Department, Yale
University, New Haven, CT 06511}

\begin{abstract}
Experiment 865 at the Brookhaven AGS has observed the decay
$K^+\rightarrow e^+\nu_e\mu^+\mu^-$. The branching ratio
extracted is 
$(1.72\pm 0.37{\rm (stat)} \pm 0.17{\rm (syst)} 
\pm 0.19{\rm (model)})\times10^{-8}$ where the third term in
the error results from the use of a model to extrapolate into
a kinematic region dominated by background.
\end{abstract}


\maketitle

The internally converted, radiative $K_{l2}$ decays, 
$K^+\rightarrow l^+\nu {l'}^+{l'}^-$, are an important source of
information on the kaon. For example, within the framework of
Chiral Perturbation Theory (ChPT) \cite{ChPT} radiative kaon
decays can serve both as an important test and a source of input
parameters for the theory. 

\begin{figure}[h]
\includegraphics[width=8.5cm]{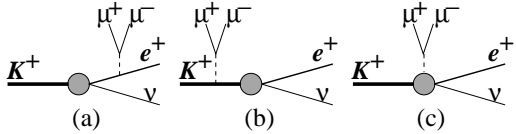}
\vspace{-0.2in}
\caption{\label{fig:graph} Graphs for contributions to
$K^+\rightarrow e^+\nu\mu^+\mu^-$.}
\end{figure}

The $K^+\rightarrow l^+\nu{l'}^+{l'}^-$ is described by the graphs
of Fig.\ \ref{fig:graph}. The tree diagrams (a) and (b) are the
inner bremsstrahlung (IB), where the virtual photon is radiated by
the kaon or positron. This contribution is electron-helicity
suppressed and is negligible for the decay 
$K^+\rightarrow e^+\nu\mu^+\mu^-$. The short distance,
structure-dependent (SD) terms are combined
in graph (c). The SD contribution is characterised by form
factors $F_V$, $F_A$, and $R$, which we define in accordance
with the Particle Data Group \cite{PDG}. These may be functions of
$W^2$ and $q^2$, the invariant masses of the $e^+\nu$ and the
$\mu^+\mu^-$ pairs, respectively.  In the vector meson dominance
picture \cite{VMD}, this dependence has the form
\begin{equation}
F_V(q^2,W^2)=F_V(0,0)/[(1-q^2/m_{\rho}^2)(1-W^2/m_V^2)]
\label{eq:FF}
\end{equation}
\noindent with similar expressions for $F_A$ and $R$. Here, 
$m_\rho$ is
the $\rho$-meson mass and $m_V$ is the mass of the $K^*(892)$
for $F_V$ and of the $K^*(1270)$ for $F_A$ and $R$. ChPT
relates the form factors $F_V$ and $F_A$ to those of the 
$\pi\rightarrow e\nu\gamma$ decay and form factor $R$ to the
kaon charge radius. Bijnens {\em et al.\ }\cite{Bijnens} gave
a ChPT prediction for the $K^+\rightarrow e^+\nu\mu^+\mu^-$
branching ratio of $1.12\times10^{-8}$. The previous
experimental limit was $<5.0\times10^{-7}$ \cite{PDG,787}.

Experiment E865 at the Brookhaven National Laboratory Alternating
Gradient Synchrotron (AGS) has
produced substantial improvements in our knowledge of these
radiative decays. Results for the decays 
$K^+\to \mu^+\nu e^+e^-$ and $K^+\to e^+\nu e^+e^-$ have already
been reported \cite{E865:AAP}. This paper presents results from
E865 for the first observation of the decay 
$K^+\to e^+\nu \mu^+\mu^-$.

\begin{figure}[h]
\includegraphics[width=8.4cm]{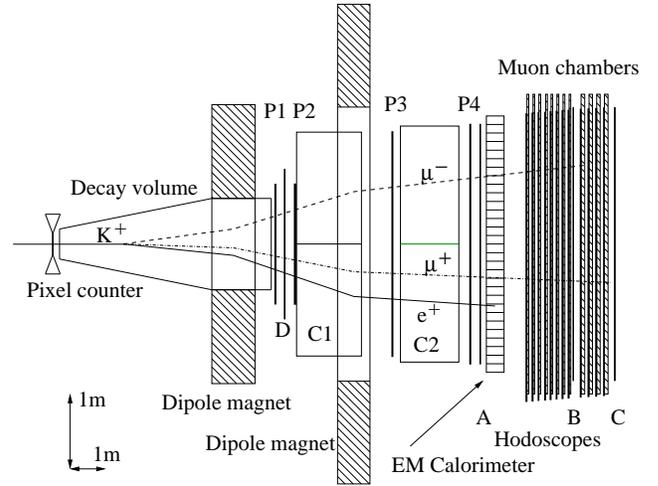}
\caption{\label{fig:app} Plan view of the experimental apparatus
for E865.}
\end{figure}

The experimental apparatus for Brookhaven E865 is shown in
Fig.\ \ref{fig:app} and has been described
in detail elsewhere \cite{E865:NIM}. A 6 GeV/c unseparated beam from
the AGS was incident on a 5-m long evacuated decay volume. Decay 
particles from this region were separated into positive and
negative charges by a dipole magnet and were momentum-analysed by a
spectrometer containing a second dipole magnet and four
4-view wire chambers, P1 -- P4. Particle identification was provided
by two pairs of gas Cherenkov counters, C1 and C2 (hydrogen on the 
negative-particle side and methane on the positive side), a 
$30\times 20$-element Shashlyk calorimeter containing 15 radiation
lengths of lead/scintillator sandwich, and a muon detector
containing twelve layers of iron with twelve 2-view wire chambers
interspersed. Additionally, there were 4 hodoscope planes, 
A -- D, for timing and triggering. A 
$12{\rm (horizontal)}\times 32{\rm (vertical)}$-element
pixel detector, with pixel size $7\times 7$ mm, was located in the
incident kaon beam to determine the position of the decaying kaon
at the entrance to the decay volume.
 
The data taking for this part of E865 took place in parallel with
studies of the $K^+\rightarrow \pi^+\pi^-e^+\nu$ ($K_{e4}$) and 
$K^+\rightarrow \pi^+\mu^+\mu^-$ decays, which have already been
published \cite{Ke4A,Ke4B,Kmumu}.

Selection of candidate events required three tracks giving a vertex 
$z$-coordinate within the decay region and an acceptable value of
$S$, where $S^2$ is the sum of squares of the deviations of the
three tracks from the fitted vertex. Also, the individual
tracks were each required to have good $\chi^2$ for the
reconstruction and good timing. Electron identification required
signals in both positive-side Cherenkov detectors and
an energy deposition in the calorimeter equal to the
reconstructed track momentum. The above cuts were also used in
the event selection for the $K_{e4}$ analysis. Additionally, for
$K^+\rightarrow e^+\nu \mu^+\mu^-$,
the muon candidate tracks were required to have sufficient hits
in the muon wire chambers, a hit in the appropriate element of
the B hodoscope, located in the middle of the muon stack, and a
signal consistent with minimum ionising in the shower calorimeter.

For each event, the neutrino momentum was calculated from the
missing momentum; ${\bf p}_{\nu}={\bf p}_{K^+}-{\bf p}_{\mu^+}-
{\bf p}_{\mu^-}-{\bf p}_{e^+}$. The magnitude of ${\bf p}_{K^+}$
was taken as the average value, derived from measurements of
pion momenta from $K_{3\pi}$ decays. The
direction of ${\bf p}_{K^+}$ was determined from the beam pixel 
detector and the reconstructed vertex where possible. About 45\%
of events had an unambiguous hit in the beam pixel detector.
For those events that did not, the average kaon beam direction
was assumed in the event reconstruction.
  
Simulation of the experiment was carried out using the GEANT
package \cite{GEANT}. A problem with this package is that pion
interactions are not always well simulated. As a result, the
probability that a pion can penetrate well into the muon stack,
causing it to be misidentified as a muon, is not well determined
by the simulation. Experimental studies using 
$K^+\rightarrow\pi^+\pi^0$, $K^+\rightarrow\pi^0\mu^+\nu$
(each followed by $\pi^0\rightarrow \gamma e^+e^-$) and
$K^+\rightarrow\pi^+\pi^+\pi^-$ events established that this $\pi$
to $\mu$ misidentification probability is calculated by the
simulation with an uncertainty of about 10\%.
 
\begin{figure}[h]
\includegraphics[width=6.4cm]{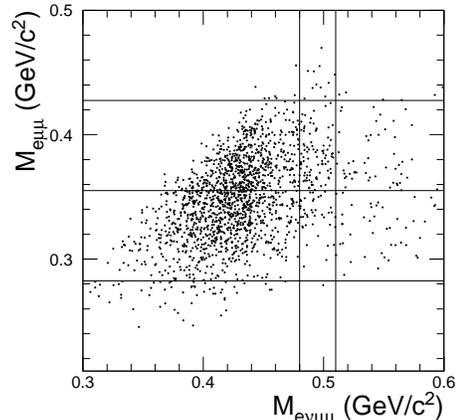}
\vspace{-0.25in}
\caption{\label{fig:data} Scatter plot of $m_{e^+\mu^+\mu^-}$
against $m_{e^+\nu\mu^+\mu^-}$ for candidate
$K^+\rightarrow e^+\nu\mu^+\mu^-$ events. Valid events lie
in the band $m_{e^+\nu\mu^+\mu^-}\sim m_K$ indicated on the plot.}
\end{figure}

With the above cuts, 1834 candidate events remain. These are 
shown in Fig.\ \ref{fig:data} as a scatter plot of 
$m_{e^+\mu^+\mu^-}$ against $m_{e^+\nu\mu^+\mu^-}$, which also
shows the region of $m_{e^+\nu\mu^+\mu^-}$ where genuine events
are expected.

Three sources of background were considered:
 
\hspace{0.1in}(a) $K^+\rightarrow\pi^+\pi^-e^+\nu$ ($K_{e4}$) with
both pions 
 
\hspace{0.4in}misidentified as muons,

\hspace{0.1in}(b) Accidentals, and

\hspace{0.1in}(c) $K^+\rightarrow\pi^+\pi^+\pi^-$ ($K_{3\pi}$) with
the pions 
 
\hspace{0.4in}misidentified as $e^+$, $\mu^+$ and $\mu^-$
respectively.
 
\noindent The contributions from these were determined as follows.
 
$K_{e4}$ events give the largest contribution to the background.
A large sample of Monte Carlo events ($6\times 10^7$, equivalent
to 20 times the number of data events) was generated. 
Background hits taken from actual data events were added to the
simulated events, and the sample was analysed as for data events.
This procedure should give a realistic estimate of the background
shape. The magnitude of he $K_{e4}$ background has an error due
to uncertainty in the $\pi$ to $\mu$ misidentification probability,
which is about 0.05 with an error of $\sim 10\%$.
 
A sample of accidental events was extracted from the data by
selecting events with bad timing and badly reconstructed vertices.
The reconstructed total charged-track momenta for these events
shows a tail above 6.5 GeV/c. The total charged-track momentum
spectrum for the 
$K^+\rightarrow e^+\nu\mu^+\mu^-$ candidates shows a similar tail,
which is assumed to arise entirely from accidentals.
Therefore, the background contribution from accidentals 
was assumed to have the
same shape as the bad-timing, bad-vertex events, with a magnitude
derived by scaling this high-momentum tail to match that in the
data. 

The background contribution from $K_{3\pi}$ decays was estimated 
by examining the $K^+\rightarrow \pi^+\pi^+\pi^-$ data events
obtained from the minimum-bias trigger. The $\mu^+$ and $\mu^-$
identification was required but there was no particle identification
requirement on the third particle. The sample was then scaled by
the known $\pi^+$ to $e^+$ misidentification probability of 
$1.3\times 10^{-3}$, determined from a study of $K_{e4}$
events, where the contamination from $K_{3\pi}$ decays is
easily identified by the kinematics of the $3\pi$ final state.
 
Background from $K^+\rightarrow\pi^+\pi^-\mu^+\nu$ is negligible
because of the small $\pi^+$ to $e^+$ misidentification
probability $(1.3\times 10^{-3})$ and the low branching
ratio for this decay.

\begin{figure}[t]
\includegraphics[width=9.0cm]{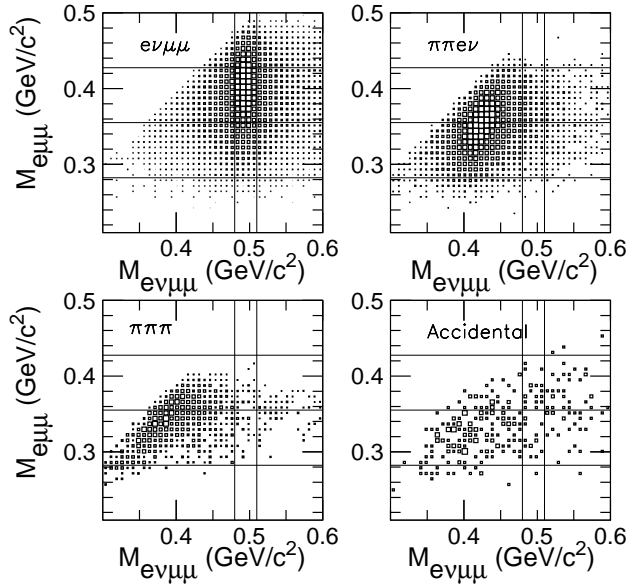}
\vspace{-0.35in}\caption{\label{fig:mc} Simulation results for the
$K^+\rightarrow e^+\nu\mu^+\mu^-$ (top left-hand plot) and
the computed contributions from the three background sources,
$K_{e4}$, $K_{3\pi}$ and accidentals.}
\end{figure}
 
The three sources of background are shown as plots of
$m_{e^+\mu^+\mu^-}$ against $m_{e^+\nu\mu^+\mu^-}$ in 
Fig.\ \ref{fig:mc}, which also shows the simulated 
$K^+\rightarrow e^+\nu\mu^+\mu^-$ signal. The signal appears
as a peak in $m_{e\nu\mu\mu}$ at the kaon mass, 
$0.494~{\rm GeV}/c^2$, the region indicated by the vertical 
lines in Figs.\ \ref{fig:data} and \ref{fig:mc}.
It is also apparent from Fig.\ \ref{fig:mc},
that the signal-to-background ratio will be best for
high values of $m_{e\mu\mu}.$ 

Because the signal-to-background ratio becomes poor at
low $m_{e\mu\mu}$, it is difficult to extract significant
information on form factors from the present experiment.
Initially, therefore, we assumed form factors for 
$K^+\rightarrow e^+\nu\mu^+\mu^-$ in order to extrapolate the 
signal into regions where it is not observable, to extract the
total branching ratio. To do so, the 2-dimensional plot, 
Fig.\ \ref{fig:data}, was fitted with the sum of the signal and
background contributions shown in Fig.\ \ref{fig:mc}. 
While the magnitude of the signal was a free parameter in
the fit, the shape of the
signal distribution was fixed by the form factors $F_A$, 
$F_V$ and $R$, whose values 
are taken from the
E865 measurements on $K^+\rightarrow e^+\nu e^+e^-$ and  
$\mu^+\nu e^+e^-$ \cite{E865:AAP}:
 \begin{eqnarray}
F_V(0,0)&=&0.112\pm 0.015{\rm (stat)}\pm 0.010{\rm (syst)}
\nonumber \\
 &~&~\pm 0.003{\rm (model)},
\label{eq:AAP:FV}
\end{eqnarray}
\begin{eqnarray}
F_A(0,0)&=&0.035\pm 0.014{\rm (stat)}\pm 0.013{\rm (syst)}
\nonumber \\
 &~&~\pm 0.003{\rm (model)},
\label{eq:AAP:FA}
\end{eqnarray}
\begin{eqnarray}
R(0,0)&=&0.227\pm 0.013{\rm (stat)}\pm 0.010{\rm (syst)}
\nonumber \\
 &~&~\pm 0.009{\rm (model)}.
\label{eq:AAP:R}
\end{eqnarray}
\noindent Since the $K_{e4}$ contribution to the background is
uncertain to about 10\%, we allow the magnitude of it to float in
the fitting. 
 
\begin{figure}[t]
\includegraphics[width=7.6cm]{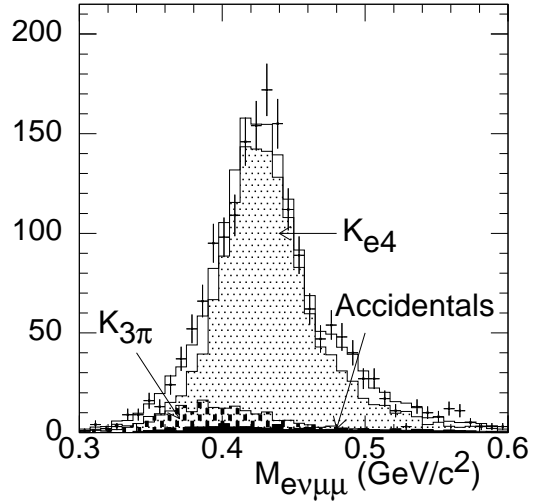}
\vspace{-0.2in}\caption{\label{fig:fit} Projection onto the 
$m_{e\nu\mu\mu}$ axis of the data of Fig.\ \ref{fig:data} 
(points) and of the fitted distribution (unshaded histogram).
The three background contributions are shown separately (shaded
histograms).}
\end{figure}
 
\begin{figure*}
\centerline{
\resizebox{.32\textwidth}{!}{
\includegraphics[width=5.5cm]{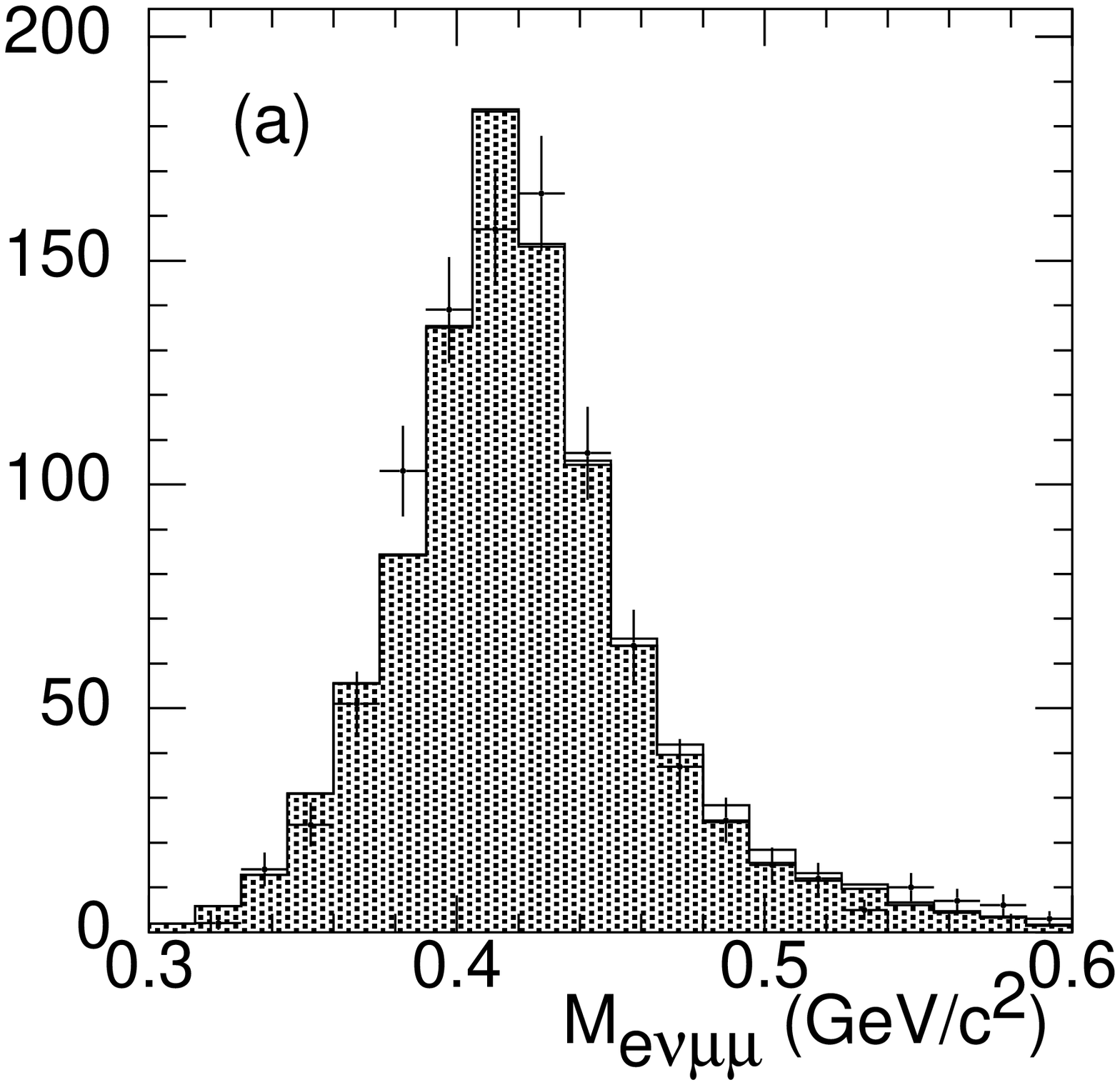}}
\hfill
\resizebox{.32\textwidth}{!}{
\includegraphics[width=5.5cm]{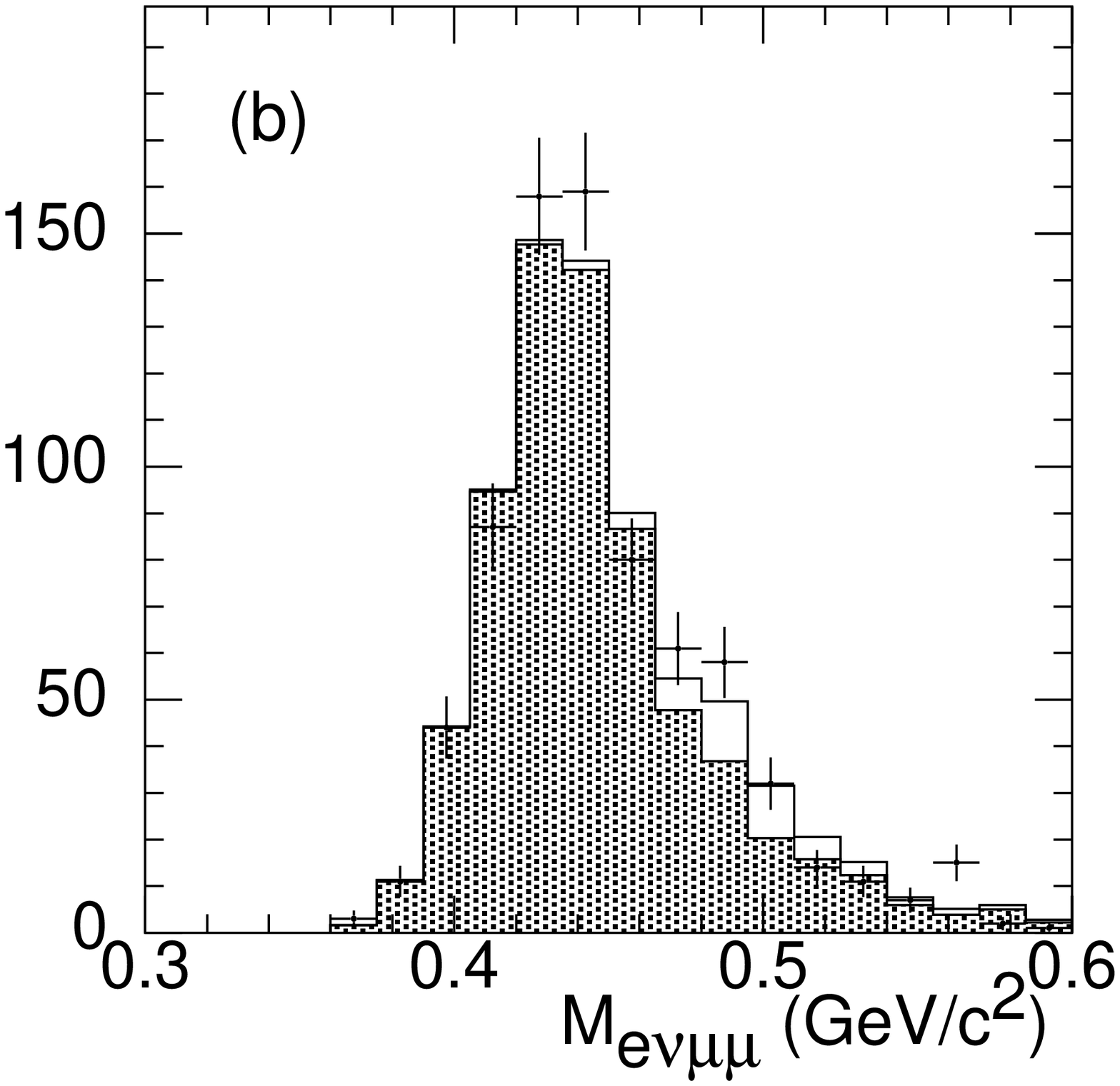}}
\hfill
\resizebox{.32\textwidth}{!}{
\includegraphics[width=5.5cm]{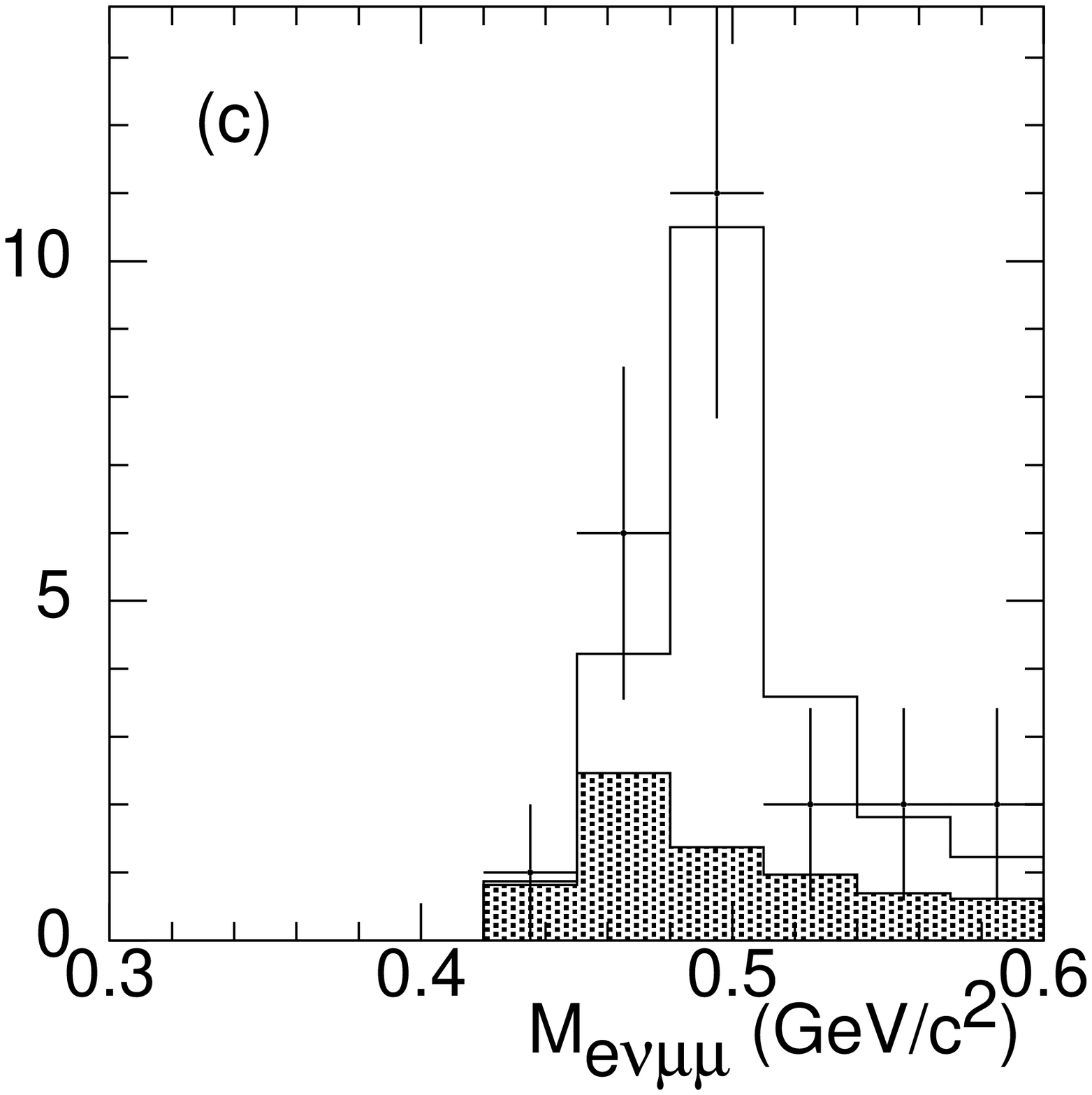}}}
\vspace{-0.2in}
\caption{\label{fig:sig} Distribution in $m_{e\nu\mu\mu}$ of
events in three bands of $m_{e\mu\mu}$: 
(a) $0.2825~{\rm GeV}/c^2 < m_{e\mu\mu} < 0.355~{\rm GeV}/c^2$,
(b) $0.355~{\rm GeV}/c^2 < m_{e\mu\mu} < 0.4275~{\rm GeV}/c^2$ and
(c) $m_{e\mu\mu} > 0.4275~{\rm GeV}/c^2$. The
points are the experimental data and the histogram is the
fitted spectrum. The shaded areas show the fitted background
contribution.}
\end{figure*}

The fit is shown in Fig.\ \ref{fig:fit} as a projection onto
the $m_{e\nu\mu\mu}$ axis of the fit and of the data from
Fig.\ \ref{fig:data}. The three contributions to the background
are shown separately. Fig.\ \ref{fig:sig} shows the same projection
for events in three bands of $m_{e\mu\mu}$ as follows:
 
Fig.\ \ref{fig:sig}(a): 
$0.2825 < m_{e\mu\mu} < 0.355~{\rm GeV}/c^2,$
 
Fig.\ \ref{fig:sig}(b): 
$0.355 < m_{e\mu\mu} < 0.4275~{\rm GeV}/c^2,$

Fig.\ \ref{fig:sig}(c): $m_{e\mu\mu} > 0.4275~{\rm GeV}/c^2.$
 
\noindent These regions are indicated by the horizontal lines
in Figs.\ \ref{fig:data} and \ref{fig:mc}. 
Fig.\ \ref{fig:sig}(c) contains 24 data events with a 
fitted background of 6.9 events, and Fig.\ \ref{fig:sig}(b)
has 176 events with a background of 133$\pm$4 events
for the reconstructed kaon mass range 
$0.465<m_{e\nu\mu\mu}<0.540~{\rm GeV/c^2}$.

The fit resulted in a branching ratio, $B$, of 
\begin{eqnarray}
B&=&(1.72\pm 0.37{\rm (stat)}\pm 0.17{\rm (form~factor)}
\nonumber \\
 &~&~\pm 0.09{\rm (slope)}\pm 0.17{\rm (syst)})
\times 10^{-8}
\label{eq:B1}
\end{eqnarray}
\noindent The $K_{e4}$ decay was used as a normalisation channel
for the analysis. Thus the systematic error arises predominantly
from uncertainties in simulation of the muon detection efficiency
since muons are not involved in $K_{e4}$. For this, we estimate
$\pm 10\%$. Additional errors arise from uncertainties in the
form factors of Eqs.\ (\ref{eq:AAP:FV}), (\ref{eq:AAP:FA}) and
(\ref{eq:AAP:R}), and also in the slopes of the $q^2$ and $W^2$
dependence of these form factors. For this contribution, we 
assume, as in Ref.\ \cite{E865:AAP}, that these slopes are 
uncertain to $\pm 30\%$. The fit gave $\chi^2=132$ for 112 degrees
of freedom. The best fit resulted in a scaling factor for the
$K_{e4}$ background of $0.86\pm 0.02$ which is consistent with
the estimated uncertainty in the $\pi$ to $\mu$ misidentification
probability of about $10\%$. Combining the errors in quadrature gives
\begin{equation}
B=(1.72\pm 0.45)\times 10^{-8}.
\label{eq:B2}
\end{equation}
A study of the statistical significance of the background
function shows that, in fitting the entire 
$m_{e\mu\mu}$ {\it vs.} $m_{e\nu\mu\mu}$
distribution, the probability of a statistical
fluctuation in the background simulating the signal is 
$<10^{-6}$.

If $F_A$, $F_V$ and $R$ are all allowed to vary in the fitting,
the resulting values for $B$, $F_A$, $F_V$ and $R$ are consistent
with Eqs.\ (\ref{eq:AAP:FV}), (\ref{eq:AAP:FA}), (\ref{eq:AAP:R})
and
(\ref{eq:B1}) but with substantially larger errors. This is the
expected consequence of the variation of signal-to-background
ratio across the range of phase space 
covered by our data. However, since $R$ is the dominant term, a 
fit was carried out with $R$ varied, with $F_A$ and $F_V$
constrained at the values of Eqs.\ (\ref{eq:AAP:FV}) and 
(\ref{eq:AAP:FA}). This fit gave
\begin{eqnarray}
R(0,0)&=&0.303\pm 0.036{\rm (stat)}\pm 0.011{\rm (form~factor)}
\nonumber \\
 &~&~\pm 0.009{\rm (slope)}\pm 0.017{\rm (syst)}.
\label{eq:B4}
\end{eqnarray}
\noindent This is consistent with Eq.\ (\ref{eq:AAP:R}) at the
1.7-standard-deviation level, and 
although our results cannot improve our knowledge
of the form factors, they can at least demonstrate consistency
with the values from the $K^+\rightarrow l^+\nu e^+e^-$ data.

The result for $B$ can be compared with the ChPT 
prediction of Bijnens {\em et al.\ }\cite{Bijnens} of 
$1.12\times 10^{-8}$. If the form factors of Eqs. 
(\ref{eq:AAP:FV}), (\ref{eq:AAP:FA}) and (\ref{eq:AAP:R}) are
used in the theoretical calculation, the prediction becomes 
$B=(1.04\pm 0.15)\times 10^{-8}$ which is consistent
with our result at the 1.5 standard-deviation level.

In summary, we have made a first observation of the decay
$K^+\rightarrow e^+\nu\mu^+\mu^-$ and have determined the branching
ratio to 25\%. We find values for the form factors that are
consistent with those for other $K^+\rightarrow l^+\nu {l'}^+{l'}^-$
decays but with less accuracy. The branching ratio is 
reasonably consistent with a ChPT prediction.

 
\begin{acknowledgments}
We gratefully acknowledge the contributions to the success
of this experiment by the staff and management of the AGS at the
Brookhaven National Laboratory, and the technical staffs of the
participating institutions.  This work was supported in part by the
U. S. Department of Energy, the National Science Foundations of the
USA, Russia and Switzerland, and the Research Corporation.
\end{acknowledgments}

\bibliography{e865enmm_5_05}

\end{document}